\documentclass[aps,nofootinbib,twocolumn]{revtex4-1}
\usepackage{epsfig}
\usepackage{color}
\usepackage[latin1]{inputenc}
\usepackage{float,amsmath}
\usepackage{graphicx}

\newcommand{\be}{\begin{eqnarray}}
\newcommand{\ee}{\end{eqnarray}}
\newcommand{\non}{\nonumber\\}
\newcommand{\bfJ}{{\bf J}}
\newcommand{\barrho}{{\bar{\rho}}}
\newcommand{\benum}{\begin{enumerate}}
\newcommand{\eenum}{\end{enumerate}}
\newcommand{\calJ}{{\cal J}}
\newcommand{\calP}{{\cal P}} 

\begin{document}

\title{3+1D hydrodynamic simulation of relativistic heavy-ion collisions}

\author{Bj\"orn Schenke}
\affiliation{Department of Physics, McGill University, 3600 University Street, Montreal, Quebec, H3A\,2T8, Canada}

\author{Sangyong Jeon}
\affiliation{Department of Physics, McGill University, 3600 University Street, Montreal, Quebec, H3A\,2T8, Canada}

\author{Charles Gale}
\affiliation{Department of Physics, McGill University, 3600 University Street, Montreal, Quebec, H3A\,2T8, Canada}

\begin{abstract}
We present \textsc{music}, an implementation of the Kurganov-Tadmor algorithm for relativistic 3+1 dimensional 
fluid dynamics in heavy-ion collision scenarios. This Riemann-solver-free, second-order, high-resolution scheme 
is characterized by a very small numerical viscosity and its ability to treat shocks and discontinuities very well.
We also incorporate a sophisticated algorithm for the determination of the freeze-out surface using a three dimensional
triangulation of the hyper-surface.
Implementing a recent lattice based equation of state, we compute $p_T$-spectra and pseudorapidity distributions
for Au+Au collisions at $\sqrt{s}=200\,{\rm GeV}$
and present results for the anisotropic flow coefficients $v_2$ and $v_4$ as a function of both $p_T$ and pseudorapidity $\eta$. 
We were able to determine $v_4$ with high numerical precision, 
finding that it does not strongly depend on the choice of initial condition or equation of state.
\end{abstract}

\maketitle


\section{Introduction}
Hydrodynamics is perhaps the simplest description of the dynamics of
a many-body system. Because it is coarse-grained, the complicated
short-distance and short-time interactions of the particles are averaged out.
Therefore, the effective degrees of freedom 
to describe the system reduce to a handful of conserved charges and their
currents instead of some multiple of the number of particles 
in the system which can be prohibitively large. 
Yet, as long as the bulk behavior of a fluid is concerned, hydrodynamics is
an indispensable and accurate tool. 

The equations of hydrodynamics are thus simple: They are just the
conservation laws and an additional equation of state 
(for dissipative hydrodynamics, constitutive relationships are also needed). 
In spite of their apparent simplicity, they can explain a vast amount of macroscopic physical phenomena
ranging from the flow of water to the flight of airplanes.
In this paper we are concerned in particular
with applying {\em ideal} hydrodynamics to 
the description of extremely hot and extremely dense fluids - 
the Quark-Gluon Plasma (QGP) and hadron gas.\footnote{%
We will report on the extension of the current
approach including viscous effects in another publication.}

The idea that ideal hydrodynamics \cite{Euler:1755} can describe the 
outcome of hadronic collisions has a long history starting from
Landau \cite{Landau:1953,Landau:1955,
Belenkij:1956cd,Carruthers:1974}. 
Subsequent developments and applications to relativistic heavy-ion collisions
have been carried out by many researchers
\cite{Bjorken:1982qr,Csernai:1982zz,Clare:1986qj,Amelin:1991cs,Srivastava:1992gh,Srivastava:1992cg,
Rischke:1995ir,Rischke:1995mt,Sollfrank:1996hd,Rischke:1996nq,Eskola:1997hz,
Dumitru:1998es,Kolb:1999it,Csernai:1999nf,Aguiar:2000hw,Kolb:2000sd,Kolb:2000fha,Nonaka:2000ek,Teaney:2000cw,Heinz:2001xi,Teaney:2001av,
Kolb:2001qz,Huovinen:2001cy,Hirano:2001eu,Muronga:2001zk,Eskola:2001bf,Hirano:2002ds,Huovinen:2002rn,Kolb:2003dz,Muronga:2003ta,
Csorgo:2003ry,Hama:2004rr,Huovinen:2003fa,Hirano:2005xf,Huovinen:2005gy,Heinz:2005bw,Eskola:2005ue,
Chaudhuri:2006jd,Baier:2006gy,Huovinen:2006jp,Muronga:2006zw,Nonaka:2006yn,Dusling:2007gi,Song:2007fn,Molnar:2008xj,
Luzum:2008cw,El:2008yy,Song:2008si,Bozek:2009ty,Peschanski:2009tg}
and continue to this day.
To describe the evolution of the system created by 
relativistic heavy-ion collisions, we need the following 5 conservation equations
\begin{eqnarray}
& \partial_\mu T^{\mu\nu} = 0\label{conservationEqns1} 
\\
& \partial_\mu J_B^\mu = 0 \label{conservationEqns2}
\end{eqnarray}
where $T^{\mu\nu}$ is the energy-momentum tensor and $J_B^\mu$ is the net baryon
current. In ideal hydrodynamics, these are usually re-expressed
using the time-like flow 4-vector $u^\mu$ as
\begin{eqnarray}
& T^{\mu\nu}_{\rm ideal} = (\varepsilon + \mathcal{P})u^\mu u^\nu - \mathcal{P} g^{\mu\nu}\,,
\\
& J_{B,\,\rm ideal}^{\mu} = \rho_{B} u^\mu \,,
\end{eqnarray}
where $\varepsilon$ is the energy density, $\mathcal{P}$ is the pressure, $\rho_B$
is the baryon density and 
$g^{\mu\nu} = \hbox{diag}(1, -1, -1, -1)$ is the metric tensor.
The equations are then closed by adding the equilibrium 
equation of state
\begin{eqnarray}
\mathcal{P} = \mathcal{P}(\varepsilon, \rho_B)
\end{eqnarray}
as a local constraint on the variables.

In a first attempt to use these equations to study the QGP
produced in relativistic heavy-ion collisions \cite{Bjorken:1982qr} it was
argued that at a very large $\sqrt{s}$ the boost invariant
approximation should work well. Therefore one can eliminate the
longitudinal direction from the full 3D space. Further, it was assumed that 
the heavy ions are large enough so that the system is uniform in the transverse
plane, thus eliminating all spatial dimensions from the equations. 
The energy-momentum conservation equation then simply becomes 
\begin{eqnarray}
{d\varepsilon\over d\tau} = -{\varepsilon + \mathcal{P}\over \tau}\,,
\label{eq:Bjorken}
\end{eqnarray}
where $\tau$ is defined as
\begin{eqnarray}
\tau = \sqrt{t^2 - z^2}\,,
\end{eqnarray}
together with the space-time rapidity $\eta_s$ which transforms $t,z$
coordinates to $\tau,\eta_s$ coordinates as follows 
\begin{eqnarray}
t &=& \tau\cosh\eta_s\,,\nonumber\\
z &=& \tau\sinh\eta_s\,.
\label{eq:tz}
\end{eqnarray}
If the equation of state is given by 
\begin{eqnarray}
\mathcal{P} = v_s^2 \varepsilon\,,
\end{eqnarray}
where $v_s^2$ is the speed of sound squared, then 
we can easily find the solution of Eq.(\ref{eq:Bjorken})
\begin{eqnarray}
\varepsilon_{\rm Bjorken} = \varepsilon_0 \left( \tau_0\over \tau
\right)^{1+v_s^2}
\end{eqnarray}
where $\varepsilon_0$ is the initial energy density at the initial time
$\tau_0$.

Although this solution
is too simple to realistically describe relativistic
heavy-ion collisions, it still is a good first approximation for the mid-rapidity
dynamics. However, when one starts to ask more detailed questions about the
dynamics of the evolving QGP such as the elliptic flow and HBT radii, it is
not enough. One needs more sophisticated calculations.
One of the first attempts to go beyond the Bjorken scenario was carried out
in \cite{Venugopalan:1991kg,Venugopalan:1994ux}. 
In the latter, the authors assumed cylindrical symmetry
but otherwise used a fully three dimensional formulation and were successful
in describing some SPS results available at the time.

At SPS energy, the central plateau in the rapidity distribution is not
very pronounced. It is more or less consistent with a Gaussian shape. 
In contrast, the central plateau extends over 4 units of rapidity at RHIC.
Hence, as long as one is concerned only with the dynamics near the mid-rapidity region, boost invariance should be a valid approximation at RHIC, 
restricting the relevant spatial dimensions to the transverse plane. 
Pioneering work on such 2+1 dimensional ideal hydrodynamics
was carried out in 
\cite{Kolb:1999it,Kolb:2000sd,Kolb:2000fha,Heinz:2001xi,Kolb:2001qz} 
and \cite{Huovinen:2001cy,Huovinen:2002rn}.
Much success has been achieved by these 2+1D calculations, in fact, too much to review 
in this work.  Interested readers are referred to \cite{Huovinen:2003fa,Kolb:2003dz}
for a thorough review and exhaustive references.  

To go beyond 2+1D ideal hydrodynamics is challenging. There are two main
ingredients that need to be added -- viscosities and the proper longitudinal dynamics. 
Both require major changes in algorithm and computing resources.
The main challenge for incorporating viscosities into the algorithm
is the appearance of the faster-than-light propagation of information.
The Israel-Steward formalism \cite{Israel:1976tn,Stewart:1977,Israel:1979wp}
avoids this super-luminal propagation \cite{Hiscock:1983zz,Hiscock:1985zz,Pu:2009fj} as does the more recent approach in 
\cite{Muronga:2001zk}. 
Since then a few groups have produced 2+1D
viscous hydrodynamic calculations. Two groups, 
\cite{Baier:2006gy,Romatschke:2007jx,Romatschke:2007mq,Luzum:2008cw} 
and \cite{Heinz:2005bw,Song:2007fn,Song:2008si}, use 
the Israel-Stewart formalism of viscous hydrodynamics whereas
another \cite{Dusling:2007gi} uses the \"Ottinger-Grmela \cite{Grmela:1997zz,Grmela:1997zy,Ottinger:1998zz} formalism. 
There is much to discuss on the formalism of viscous hydrodynamics alone,
but since it is not the main topic of this paper, we would like to defer
the detailed discussion to our next publication where we will present our
own viscous hydrodynamic calculations.

The motivation to construct 3+1D hydrodynamics is to investigate 
the non-trivial longitudinal dynamics and its effects on the rapidity 
dependence of the transverse dynamics.
Constructing a 3+1 dimensional ideal hydrodynamics code, however, 
is not as simple as just adding
one more dimension or one more equation to a code. 
The construction of a shock capturing algorithm, the freeze-out surface, all become much more
intricate. So far there have been a few groups who have published their
study of heavy-ion physics using realistic 3+1D ideal hydrodynamic simulations. 
One of them \cite{Hirano:2001yi,Hirano:2001eu,Hirano:2002ds}
uses a fixed grid (Eulerian) algorithm to solve the hydrodynamic equations, 
another \cite{Nonaka:2000ek,Nonaka:2006yn} 
uses a Lagrangian approach which follows the evolution of each fluid cell. 
A somewhat different approach called smoothed particle hydrodynamics is used in Ref. \cite{Aguiar:2000hw}. 

We have three major motivations to add another implementation to this list. 
The parameters such as the initial temperature profile and expansion rate
can differ between the 2+1D calculations and the 3+1D calculations and
also among different approaches. Intuitively, it is clear that
reality should favour 3+1D hydrodynamics. But since there are so many
unknowns in the initial state, such as the exact initial energy density profile,
initial flow profile and the initial baryon density profile, having an
independent algorithm is important for verifying our understanding
of the initial condition and its uncertainty. This difference in 2+1D and
3+1D initial conditions is also important in jet quenching calculations
since initial conditions can make a fair amount of difference 
in fixing the jet quenching parameters such as the initial temperature and
the effective coupling constant.

Another motivation is the desire to have a modular hydrodynamics code to
which we can couple a high $p_T$ jet physics model such as in \textsc{martini} \cite{Schenke:2009gb}.
This is to examine the response of the medium to the propagating jet as 
it loses energy to its surrounding medium. We are not yet at this stage but
planning on implementing it in the near future.

Last but not least, one should take advantage of 
recent progress in shock capturing algorithms
to possibly simplify and certainly improve the calculations, creating an updated standard from which to assess the importance of viscous effects. 
The algorithm we use is usually referred to 
as Kurganov-Tadmor method \cite{Kurganov:2000}. 

In this work we first review the Kurganov-Tadmor scheme (Section \ref{kt})
and present the implementation for relativistic ideal hydrodynamics
in a three dimensional expanding geometry (Section \ref{implementation}). 
After discussing initial conditions (Section \ref{ini}) and the employed equations of state, which
include a recent parametrization of a combined lattice and hadron resonance gas equation of state (Section \ref{eos}), 
we introduce a new algorithm for determining 
the freeze-out surface by discretizing the three dimensional hyper-surface into tetrahedra (Section \ref{freeze}).
Finally we show first results for particle spectra including resonance decays from resonances up to $2\,{\rm GeV}$,
elliptic flow, and the anisotropic flow coefficient $v_4$ (Section \ref{results}).
It is demonstrated that the latter is highly sensitive to discretization errors which are shown to be well under control
for fine enough lattices.

\section{Kurganov-Tadmor Method}
\label{kt}
Hydrodynamic equations stem from conservation laws. Hence, 
they take the following general form:
\be
\partial_t \rho_a = -\nabla{\cdot}\bfJ_a\,,
\label{eq:conserv_a}
\ee
where $a$ runs from 0 to 4, labelling the energy, 3 components of the momentum and the net baryon
density. 
The task is then
to solve these equations together with the equation of state.
Even though they 
have a deceptively simple form, they are remarkably subtle to solve.
In this section, we briefly sketch the Kurganov and Tadmor scheme (KT) \cite{Kurganov:2000}, which we use for the solution of Eq.(\ref{eq:conserv_a}). 

To illustrate the method,
consider the following single component conservation
equation in 1 spatial dimension
\be
\partial_t \rho = -\partial_x J\,,
\label{eq:conserv_1}
\ee
together with an equation that relates $J$ to $\rho$ such as $J = v \rho$.
All the essential features of KT can be explained with
this simple example. As it was shown in Ref.\,\cite{Kurganov:2000}, 
higher dimensions can be dealt with by simply
repeating the treatment here for all spatial dimensions.
Coupled conservation equations make the calculation of the maximum
local propagation speed more complicated (see below),
but there is no conceptual complication in doing so.

The need for more sophisticated numerical methods in solving conservative
equations in part comes from the fact
that a naive discretization of Eq.\,(\ref{eq:conserv_1}) such as 
\be
{\rho^{n+1}_{j}-\rho^{n}_{j}\over \Delta t}
=
-{J^{n}_{j+1}-J^{n}_{j-1}\over 2\Delta x}\,,
\label{eq:naive}
\ee
with $J = v\rho$ is unconditionally unstable. That is, the solution will
either grow without bound as $t$ increases or start to oscillate
uncontrollably.
Here the superscript $n$ indicates that the quantity is evaluated at 
$t_n=t_0 + n\Delta t$ and the subscript $j$ indicates that the quantity
represents the value at $x_j = j\Delta x$. 
One can make this stable if one devises a scheme where numerical damping is
introduced. For instance, 
suppose one replaces $\rho_j^n$ in the left hand side of Eq.\,(\ref{eq:naive})
with the spatial average $(\rho_{j+1}^n + \rho_{j-1}^n)/2$. 
In the small $\Delta t$ and $\Delta x$ limit, 
this well-known Lax method \cite{Lax:1954,Friedrichs:1954} is equivalent to solving
\be
\partial_t \rho = -\partial_x J + \left( {(\Delta x)^2\over 2\Delta t}\right)
\partial_x^2 \rho\,.
\ee
The second term is the
numerical dissipation term often referred to as the ``numerical viscosity''.
Different schemes introduce different forms of the numerical viscosity 
term. 

This simple method does stabilize the numerical solutions but 
one also can immediately see that 
$(\Delta x)^2/\Delta t$ must not be large. Otherwise, this artificial term
will dominate the numerical evolution of the system. 
Therefore in this method,
a finer time resolution result cannot be computed 
without making the number of spatial grid points correspondingly large.
Many other numerical methods also
have a $1/\Delta t$ behavior for the artificial viscosity, including KT's 
immediate predecessor \cite{Nessyahu:1990}.
However, in KT the artificial viscosity 
{\em does not} depend on $\Delta t$. It only depends on some positive power
of $\Delta x$ and we are free to take the $\Delta t\to 0$ limit.
As a bonus, this allows us to turn a set of difference equations into a set of
ordinary differential equations as explained below. 
This places the vast array of ODE solvers at one's disposal,
thus making this method much more versatile.

Notably, KT is a MUSCL-type
(Monotonic Upstream-centered Schemes for Conservation Laws) 
finite volume method \cite{Leer}
in which the cell average of the density $\rho$ around $x_j$ is used
instead of the value of the density at $x_j$.
Then, the conservation equation for the cell average
\be
\barrho_j(t) = 
{1\over \Delta x} \int_{x_{j-1/2}}^{x_{j+1/2}} dx\, \rho(x,t)\,,
\ee
becomes
\be
{d\over dt}\barrho_j(t) = {J(x_{j-1/2},t) - J(x_{j+1/2},t)\over \Delta x}\,,
\label{eq:finite_v1}
\ee
and the current and charge density at values other than the $x_j$ are contructed using
a piecewise linear approximation.
This method leads to discontinuities at the halfway points $x_{j\pm 1/2}$ where the current 
is evaluated.
Kurganov and Tadmor solved this problem using the maximal local propagation speed 
$a = \left| \partial J/\partial \rho \right|$ to identify 
how far the influence of the discontinuities at $x_{j\pm 1/2}$ could travel, and
divided the space into two groups; one with elements that
include a discontinuity and one where the solution is smooth.
The exact procedure of doing this is explained in Appendix \ref{app:kt}.

Here, we quote Kurganov and Tadmor's final result for the conservation equation in the $\Delta t\to 0$ limit:
\be
{d\over dt}\barrho_j(t) = -{H_{j+1/2}(t) - H_{j-1/2}(t)\over \Delta x}\,,
\ee
where 
\be
H_{j\pm 1/2} 
&=&
{J(x_{j\pm 1/2,+},t) + J(x_{j\pm 1/2,-},t)\over 2}\non & & {}
\hspace{-0.4cm}-
{a_{j\pm 1/2}(t)\over 2}
\left(
\barrho_{j\pm 1/2,+}(t) - \barrho_{j\pm 1/2,-}(t)
\right)\,,
\ee
with
\be
\barrho_{j+1/2,+} &=& \barrho_{j+1}-{\Delta x\over 2}(\rho_x)_{j+1}\,,
\\
\barrho_{j+1/2,-} &=& \barrho_j + {\Delta x\over 2}(\rho_x)_j\,.
\ee

The order of the spatial derivatives $(\rho_x)_j$ is chosen by the minmod flux limiter
\begin{equation}
(\rho_x)_j
=
\hbox{minmod}
\left(
\theta {\barrho_{j+1}-\barrho_j\over \Delta x}, 
{\barrho_{j+1}-\barrho_{j-1}\over 2\Delta x},
\theta {\barrho_{j}-\barrho_{j-1}\over \Delta x}
\right)\nonumber
\end{equation}
where
\be
\hbox{minmod}(x_1, x_2, \cdots) = \left\{
\begin{array}{ll}
\hbox{min}_j\{x_j\},& \hbox{if $x_j> 0$ $\forall j$}\\
\hbox{max}_j\{x_j\},& \hbox{if $x_j< 0$ $\forall j$}\\
0, & \hbox{otherwise}
\end{array}
\right.\nonumber
\ee
and $1\le \theta \le 2$ is a parameter that controls the amount of
diffusion and the oscillatory behavior. This is also our choice with $\theta
= 1.1$.
This allows for higher accuracy using the second-order approximation where possible
and avoids spurious oscillations around stiff gradients by switching to 
the first order approximation where necessary.

The Kurganov-Tadmor method, combined with a suitable flux limiter such as the one just described,  
is a non-oscillatory and simple central difference scheme 
with a small artificial viscosity which can also handle shocks very well
(for an extensive comparison with other schemes in this regard, see 
Ref.\,\cite{LucasSerrano:2004aq}). It is also Riemann-solver free and hence
does not require calculating the local characteristics. 
This scheme is ideally suited for hydrodynamics studies.

\section{Implementation}
\label{implementation}\hyphenation{music}
 We now describe our implementation of the KT algorithm for relativistic heavy-ion collisions, 
 dubbed \textsc{music}, MUScl for Ion Collisions.

 As in most ideal hydrodynamics implementations for heavy-ion collisions, 
 the most natural coordinate system for us is the $\tau-\eta_s$ coordinate
 system defined by Eq.(\ref{eq:tz}).
 In the $\tau-\eta_s$ coordinate system, the conservation equation
 $\partial_\mu J^\mu = 0$ becomes
 \be
  \partial_\tau (\tau J^\tau)
 +
 \partial_{\eta_s} J^{\eta_s}
 + \partial_v(\tau J^v)& = &0\,,
 \label{eq:jcons}
 \ee
 where
 \be
 J^\tau &= & (\cosh\eta_s J^0 - \sinh\eta_s J^3)\,,
 \label{eq:jtau}
 \\
 J^{\eta_s} &= &
 ( \cosh\eta_s J^3 - \sinh\eta_s J^0 )\,,
 \label{eq:jeta}
 \ee
 which is nothing but a Lorentz boost with the space-time rapidity
 $\eta_s = \tanh^{-1}(z/t)$. 
 The index $v$ and $w$ in this
 section always refer to the transverse $x,y$ coordinates which are not
 affected by the boost.
 Applying the same transformation to both indices of $T^{\mu\nu}$,
 one obtains
\be
\partial_\tau (\tau T^{\tau\tau})
+ \partial_{\eta_s} (T^{\eta_s\tau}) + \partial_v (\tau T^{v\tau})
+ T^{\eta_s\eta_s} = 0\,,
\label{eq:econs}
\ee
and
\be
\partial_\tau (\tau T^{\tau\eta_s})
+ \partial_{\eta_s}(T^{\eta_s\eta_s})
+ \partial_v(\tau T^{v\eta_s}) + T^{\tau\eta_s} = 0\,,
\label{eq:etacons}
\ee
and
\be
\partial_\tau (\tau T^{\tau v})
+ 
\partial_{\eta_s} (T^{\eta_s v})
+
\partial_w (\tau T^{w v})
& = &
0\,,
\label{eq:xycons}
\ee
These 5 equations, namely Eq.\,(\ref{eq:jcons}) for the net baryon current, and
Eqs.\,(\ref{eq:econs}, \ref{eq:etacons}, \ref{eq:xycons}) for the energy and
momentum are the equations we solve with the KT scheme
explained in the previous section.
Multi-dimension is dealt with by repeating the KT scheme in each direction
\cite{Kurganov:2000}.
The source term is dealt with by following the suggestions in the original
KT paper and others \cite{Naidoo:2004}.

At each time step, the new values of $J^\tau,T^{\tau\tau},T^{\tau\eta_s}$ and
$T^{\tau v}$ are obtained by solving the semi-discrete version of KT using Heun's rule.
Heun's rule is a form of the second-order
Runge-Kutta method which can be stated as follows. 
Suppose we have a differential equation 
\be
{d\rho\over dt} = f(t,\rho)\,.
\ee
A numerical solution of this equation can be obtained by applying
the following rules
\benum
 \item Compute $k_1 = f(t,\rho_n)$.
 \item Compute $\rho_{n+1}' = \rho_n + k_1\Delta t$.
 \item Compute $k_2 = f(t+\Delta t, \rho_{n+1}')$.
 \item Compute $\rho_{n+1} = \rho_n + (k_1 + k_2)\Delta t/2$.
\eenum
Once new values of $J^\tau,T^{\tau\tau},T^{\eta_s\tau}$ and
$T^{v\tau}$ are obtained, 
the following ideal gas expressions
 \be
T^{\tau \tau} & = & (\varepsilon +\calP) u^\tau u^\tau - \calP
\\
T^{\tau \eta_s} & = & (\varepsilon + \calP) u^{\eta_s} u^\tau
\\ 
T^{\tau v} &=& (\varepsilon + \calP) u^\tau u^v
\\
J^\tau  & = & \rho u^\tau
\ee
together with the equation of state
\be
\mathcal{P} = \mathcal{P}(\varepsilon, \rho)
\ee
determine the net baryon density $\rho$, the pressure $\mathcal{P}$,
the energy density $\varepsilon$, and the flow velocity $u^\mu$.
The flow components $u^\tau$ and $u^{\eta_s}$ here are given by the Lorentz boost
with the space time rapidity $\eta_s$ exactly as in Eqs.\,(\ref{eq:jtau},
\ref{eq:jeta}).
Hence, they still satisfy the normalization condition 
$u_\tau^2 = 1 + u_x^2 + u_y^2 + (1/\tau^2) u_{\eta_s}^2$.

Explicitly,
the values of $\varepsilon$ and $\rho$ are obtained by iteratively solving
the following coupled equations
\be
\varepsilon 
& = &
T^{\tau\tau}
-
{K \over (T^{\tau\tau} + \mathcal{P}(\varepsilon,\rho) )}\,,
\\
\rho
& = &
{J^\tau}
\sqrt{
{\varepsilon+\mathcal{P}(\varepsilon,\rho)
\over
T^{\tau\tau}+\mathcal{P}(\varepsilon,\rho)}\,,
}
\ee
where $K = (T^{\eta_s\tau})^2 + (T^{x\tau})^2 + (T^{y\tau})^2$. 
A good initial guess turned out to be either the value 
at the previous time step or just the initial $T^{\tau\tau}$ and $J^\tau$.
Knowing $\varepsilon,\rho$, we can calculate the pressure
$\mathcal{P} = \mathcal{P}(\varepsilon,\rho)$ and $u^\tau = J^\tau/\rho$.
These then determine the spatial flow vector components as
\be
u^{i} = {T^{\tau i}\over (\varepsilon + \mathcal{P}) u^\tau}\,,
\ee
for $i=\eta_s, x, y$. 
With these $\varepsilon, \rho, \mathcal{P}$ and $u^\mu$, the whole $T^{\mu\nu}$ 
can be reconstructed and be used at the next time step.

In addition to the currents,
we need to find the maxi\-mum local propagation speed at each time step.
The maximum speed in the $k$ direction
is given by the maximum eigenvalue of the following Jacobian:
\be
\calJ^k_{ab} = {\partial J^{k}_a\over \partial J^{\tau}_b}
\ee
where $J_a^\mu$ with $a=0,1,\cdots,4$ stand for the 5 currents (net baryon,
energy and momentum).
The whole matrix is quite complicated.
However, with the help of \textsc{mathematica} \cite{Math}, 
it turned out that the eigenvalues can be analytically calculated.

If there is no net baryon current to consider, 
two of the 4 eigenvalues  
in the $k = x,y$ direction are degenerate and equal to
$u^k/u^\tau$. The remaining two are
\be
\lambda_k^{\pm} &=&
\frac{A\pm\sqrt{B}}{D }\,,
\ee
with 
\be
A&=&u^\tau u^k(1-v_s^2)\,,\non 
B&=&\left[ u_\tau^2-u_k^2 -\left(u_\tau^2-u_k^2-1\right) 
v_s^2 \right]v_s^2\,,\non
D&=& u_\tau^2 -\left(u_\tau^2-1\right) v_s^2\,.
\label{eq:ABD}
\ee
where $v_s^2 = \calP'(\varepsilon)$ is the speed of sound squared.
In the $k=\eta_s$ direction,
we have the same expression, but the eigenvalues are scaled with $1/\tau$,
that is $\lambda_{\eta_s} = \lambda_{k\to\eta_s}/\tau$. 
The same expressions for the Cartesian case was obtained in \cite{Banyuls:1997}.
The largest eigenvalue is thus
\be
|\lambda_k^{\rm largest}|
&=&
\frac{
|A|+\sqrt{B}}{ D }\,,
\label{eq:largest}
\ee
with an additional $1/\tau$ factor for $k=\eta_s$.

Even with the net baryon current present, 
the eigenvalues of the the resulting $5\times 5$ matrix
can be computed analytically.
Consider first $k=x,y$ directions. Among the 5 eigenvalues, 3 are degenerate
and equal to $u^k/u^\tau$. The other two are again given by
Eq.(\ref{eq:ABD})
with 
\be
v_s^2 =
\partial_\varepsilon \calP + (\rho/(\varepsilon+\calP))\partial_\rho \calP\,.
\ee
It is obvious that the $\rho\to 0$ limit coincides with the no current case.
In the $\eta_s$ direction, the expressions for the eigenvalues are the same
except for the overall scale factor $1/\tau$.
The maximum eigenvalue is again given by Eq.(\ref{eq:largest})
with the above substitution.

Importantly, \textsc{music} is fully parallelized to run on many processors simultaneously. To achieve this,
the lattice is truncated in the $\eta_s$ direction so that each processor only has to evolve
the system on one slice, communicating the cell values at the boundary to the neighboring processors every time step.
This leads to an increase in speed almost (minus the time necessary for communication between the processors)
linear in the number of processors used.

The typical size of a time step is of the order of $0.01\,{\rm fm}/c$.
Energy conservation is fulfilled to better than 1 part in 30,000 per time step.

\section{Initial conditions}
\label{ini}
The initialization of the energy density is done using the Glauber model (see \cite{Miller:2007ri} and references therein):
Before the collision the density distribution of the two nuclei is described by
a Woods-Saxon parametrization
\begin{equation}
  \rho_A(r) = \frac{\rho_0}{1+\exp[(r-R)/d]}\,,
\end{equation}
with $R=6.38\,{\rm fm}$ and $d=0.535\,{\rm fm}$ for Au nuclei.
The normalization factor $\rho_0$ is set to fulfill $\int d^3r \rho_A(r)=A$. 
With the above parameters we get $\rho_0=0.17\,{\rm fm}^{-3}$.
The relevant quantity for the following considerations is the \emph{nuclear thickness function}
\begin{equation}
  T_A(x,y) = \int_{-\infty}^{\infty}dz\,\rho_A(x,y,z)\,,
\end{equation}
where $r=\sqrt{x^2+y^2+z^2}$.
The opacity of the nucleus is obtained by multiplying the thickness function with the total inelastic cross-section $\sigma_0$ of
a nucleus-nucleus collision.

Experiments at SPS found that the number of final state particles scales with the number of wounded
nucleons, nucleons that interact at least once in the collision. 
Deviations from the scaling are observed at RHIC.

Statistical considerations allow to express the number of wounded nucleons in the transverse plane 
by the nuclear thickness function of one nucleus, multiplied with a combinatorial factor involving the nuclear thickness function
of its collision partner. This factor ensures that the participating nucleon does not penetrate the finite opposing nuclear matter
without interaction.
For noncentral collisions of nuclei with mass numbers $A$ and $B$ at impact parameter $b$, the number of wounded nucleons per transverse
area is given by \cite{Kolb:2001qz}
\begin{align}
  &n_{\rm WN}(x,y,b) = \nonumber \\
  & T_A(x+\frac{b}{2},y)\left[1-\left(1-\frac{\sigma_0 T_B(x-\frac{b}{2},y)}{B}\right)^B\right]\nonumber\\
  &~~+ T_B(x-\frac{b}{2},y)\left[1-\left(1-\frac{\sigma_0 T_A(x+\frac{b}{2},y)}{A}\right)^A\right]\,.
\end{align}
Integrating this expression over the transverse plane yields the total number of wounded nucleons (participants)
as a function of the impact parameter. We compute the relevant quantities using routines adapted from \textsc{lexus} \cite{Jeon:1997bp}.

At high energies the density of binary collisions becomes of interest. After suffering their first collision, the partons
travel on through the nuclear medium and are eligible for further (hard) collisions with other partons.
This leads to the notion that one has to count the binary collisions. 
The density of their occurence in the transverse plane is simply expressed by the product of the thickness function of one nucleus
with the encountered opacity of the other nucleus, leading to 
\begin{equation}
  n_{\rm BC}(x,y,b) = \sigma_0 T_A(x+b/2,y)T_B(x-b/2,y)\,.
\end{equation}
The total number of binary collisions shows a stronger dependence on the impact parameter than does the number of wounded nucleons.

We now assume that the initial state of matter in the transverse plane is governed entirely by the physics of `soft' and `hard' 
processes represented in terms of the densities of wounded nucleons and binary collisions, respectively.
Shadowing effects by the spectators do not play a role at RHIC energies because the spectators leave the transverse plane at $z=0$
on a timescale of less than $1\,{\rm fm}/c$.

Whether the deposited energy density or entropy density scales with the density of wounded nucleons or binary collisions is not
clear from first principles. As mentioned above, SPS data suggests that the final state particle multiplicity is proportional to the number of wounded 
nucleons. At RHIC energies a violation of this scaling was found (the particle production per wounded nucleon is a function increasing 
with centrality. This is attributed to a significant contribution from hard processes, scaling with the number of binary collisions).

We parametrize the shape of the initial energy density distribution in the transverse plane as
\begin{equation}\label{inieps}
  W(x,y,b)=(1-\alpha)\, n_{\rm WN}(x,y,b)+\alpha\, n_{\rm BC}(x,y,b)\,,
\end{equation}
where $\alpha$ determines the fraction of the contribution from binary collisions.

Alternatively, we can scale the entropy density as opposed to the energy density as in (\ref{inieps}).
This leads to more pronounced maxima in the energy density distributions because 
of the relation $\varepsilon\sim s^{4/3}$ in the QGP phase. However, a similar effect can be achieved by increasing the contribution of
binary collision scaling $\alpha$. 

For the longitudinal profile we employ the prescription used in 
\cite{Ishii:1992xi,Morita:1999vj,Hirano:2001yi,Hirano:2001eu,Hirano:2002ds,Morita:2002av,Nonaka:2006yn}.
It is composed of two parts, a flat region around $\eta_s=0$ and half a Gaussian in the forward
and backward direction:
\begin{equation}
  H(\eta_s)=\exp\left[-\frac{(|\eta_s|-\eta_{\rm flat}/2)^2}{2 \sigma_{\eta}^2}\theta(|\eta_s|-\eta_{\rm flat}/2)\right]\,.
\end{equation}

The full energy density distribution is then given by
\begin{equation}
  \varepsilon(x,y,\eta_s,b) = \varepsilon_0\, \, H(\eta_s)\,W(x,y,b)/W(0,0,0)\,.
\end{equation} 

The parameters $\eta_{\rm flat}$ and $\sigma_\eta$ are tuned to data and will be quoted below.

\section{Equation of state}
\label{eos}
To close the set of equations (\ref{eq:jcons}, \ref{eq:econs}, \ref{eq:etacons}, \ref{eq:xycons}) we must provide a nuclear equation of 
state $\calP(\varepsilon,\rho)$
which relates the local thermodynamic quantities.
We present calculations using a modeled equation of state (EOS-Q) also used in \textsc{azhydro} \cite{Kolb:2000sd,Kolb:2002ve,Kolb:2003dz}
as well as one extracted from recent lattice QCD calculations \cite{Huovinen:2009yb}.

For the EOS-Q, the low temperature regime
is described as a non-interacting gas of hadronic resonances,
summing over all resonance states up to 2\,{\rm GeV} \cite{Amsler:2008zzb}.
Above the critical temperature $T_{\rm crit}{\,=\,}164\,{\rm MeV}$, the system is modeled as a non-interacting 
gas of massless $u$, $d$, $s$ quarks and gluons, subject to an external bag pressure $B$.
The two regimes are matched by a Maxwell construction, 
adjusting the bag constant  $B^{1/4}{\,=\,}230\,{\rm MeV}$ such that for a system 
with zero net baryon density the transition temperature coincides with 
lattice QCD results \cite{Engels:1981qx,Karsch:2001vs}.
The Maxwell construction inevitably leads to a strong first order
transition, with a large latent heat.

However, lattice results suggest a smoother transition. 
Recently, in \cite{Huovinen:2009yb} several parametrizations of the
equation of state which interpolate between the lattice data at high temperature and a hadron-resonance gas
in the low temperature region were constructed. 
We adopt the parametrization ``s95p-v1'' (and call it EOS-L in the following), where
the fit to the lattice data was done above $T = 250\,{\rm  MeV}$, 
and the entropy density was constrained at $T = 800\,{\rm  MeV}$ to be 
95\% of the Stefan-Boltzmann value. Furthermore, one "datapoint" was added to the fit to make the peak in the trace anomaly higher. 
See \cite{Huovinen:2009yb} for more details on this parametrization of the nuclear equation of state.

\section{Freeze-out}
\label{freeze}
The spectrum of produced hadrons of species $i$ with degeneracy $g_i$ is given by the Cooper-Frye formula \cite{Cooper:1974mv}:
\begin{equation}\label{cf}
E\frac{dN}{d^3p}=\frac{dN}{dy p_T dp_T d\phi_p} = g_i \int_\Sigma f(u^\mu p_\mu) p^\mu d^3\Sigma_\mu\,,
\end{equation}
with the distribution function
\begin{equation}
  f(u^\mu p_\mu) = \frac{1}{(2\pi)^3}\frac{1}{\exp((u^\mu p_\mu -\mu_i)/T_{\rm FO})\pm 1}\,.
\end{equation}
We assumed that at freeze-out every infinitesimal part of the hyper-surface 
$\Sigma$ behaves like a simple black body source of particles (this assumption will be modified when including viscosity).
The collective velocity of the fluid on the hyper-surface, which results from longitudinal and transverse flow, is taken into account
by using the invariant expression $E=E(x)=u^\mu(x) p_\mu$.
To evaluate the right hand side of (\ref{cf}) we need to determine the freeze-out hyper-surface 
\begin{align}\label{eq:Sigma}
  \Sigma&=(\Sigma^0(x,y,\eta_s),\Sigma^1(x,y,\eta_s),\Sigma^2(x,y,\eta_s),\Sigma^3(x,y,\eta_s))\nonumber\\
  &=(\tau_f(x,y,\eta_s)\cosh\eta_s,x,y,\tau_f(x,y,\eta_s)\sinh\eta_s)\,,
\end{align}
where $\tau_f(x,y,\eta_s)$ is the freeze-out time, determined by when the energy density (or temperature) falls below the critical value $\varepsilon_{\rm FO}$ (or $T_{\rm FO}$).
The normal vector on this surface is given by
\begin{equation}
  d^3\Sigma_\mu=-\varepsilon_{\mu\nu\lambda\rho} \frac{\partial\Sigma^\nu}{\partial x}\frac{\partial\Sigma^\lambda}
  {\partial y}\frac{\partial\Sigma^\rho}{\partial \eta_s}\,dxdyd\eta_s\,,
\end{equation}
with the totally anti-symmetric tensor of fourth order
\begin{equation}
  \varepsilon^{\mu\nu\lambda\rho}=-\varepsilon_{\mu\nu\lambda\rho}= \left\{ \begin{array}{rrr}
         1 & \mbox{even permutation}\\
        -1 & \mbox{odd permutation}\\
         0 & \mbox{otherwise}\end{array} \right.
\end{equation}
Using (\ref{eq:Sigma}) we find
\begin{align}
  d^3\Sigma_\mu=\Big(&\frac{\partial\tau_f}{d\eta_s}\sinh{\eta_s}+\tau_f\cosh\eta_s,
    -\tau_f\frac{\partial\tau_f}{\partial x},
    -\tau_f\frac{\partial \tau_f}{\partial y},\nonumber\\
    &-\frac{\partial \tau_f}{\partial \eta_s}\cosh\eta_s-\tau_f \sinh\eta_s\Big)dxdyd\eta_s\,.
\end{align}
To evaluate Eq.\,(\ref{cf}) we need to determine $u^\mu p_\mu$ and $p^\mu d\Sigma_\mu$.
The hydrodynamic evolution calculation provides $u^\tau, u^x, u^y, u^{\eta_s}$, 
so we express $p^\mu$ as\footnote{In this section, our definition of 
the 4-vector component $v^{\eta_s} = (\cosh\eta_s\,v^3 {-} \sinh\eta_s\,v^0)/\tau$
carries an extra factor of $1/\tau$.} 
\begin{align}
  p^\tau&=m_T \cosh(y-\eta_s)\nonumber\\
  p^{\eta_s}&=\frac{m_T}{\tau} \sinh(y-\eta_s)\,,\nonumber
\end{align}
with $m_T=\sqrt{m^2+p_T^2}$, where $m$ is the mass of the considered particle,
and obtain
\begin{align}
  u^\mu p_\mu &= u^\tau p^\tau - u^x p^x - u^y p^y 
  - \tau^2 u^{\eta_s} p^{\eta_s} \nonumber\\
             &= u^\tau m_T\cosh(y-\eta_s) - u^x p^x - u^y p^y \nonumber \\
             &~~~~ - \tau u^{\eta_s} m_T \sinh(y-\eta_s)\,.
\end{align}
Note that y's appearing in the cosh and sinh functions represent
the rapidity of the produced hadron.
We can express $\Sigma$ from Eq.\,(\ref{eq:Sigma}) in terms of $\tau-\eta_s$ coordinates
\begin{equation}
\Sigma^\alpha=(\Sigma^\tau,\Sigma^x,\Sigma^y,\Sigma^{\eta_s})=(\tau_f(x,y,\eta_s),x,y,\eta_s)\,,
\end{equation}
and get
\begin{align}\label{d3Sigma}
d^3\Sigma_\alpha &= (1,-\frac{\partial \tau_f}{\partial x},-\frac{\partial \tau_f}{\partial y},-\frac{\partial \tau_f}{\partial \eta_s})
  \sqrt{-{\rm det}\,g}\, dx dy d\eta_s\nonumber\\
 &= (1,-\frac{\partial \tau_f}{\partial x},-\frac{\partial \tau_f}{\partial y},-\frac{\partial \tau_f}{\partial \eta_s})
  \tau_f\, dx dy d\eta_s\,,
\end{align}
with $g$ the metric in $\tau-\eta_s$ coordinates.
The scalar product of (\ref{d3Sigma}) with $p^\alpha$ is then found to be
\begin{align}\label{pSigma2}
  p^\alpha &d^3\Sigma_\alpha = \nonumber\\
  &\Big[m_T\frac{\partial}{\partial\eta_s}\left(\tau_f \sinh (y-\eta_s)\right)
  - \tau_f \vec{p}_T\cdot\vec{\nabla}_T\tau_f\Big] dx dy d\eta_s\,,
\end{align}
with the two-dimensional derivative $\vec{\nabla}_T=\left(\partial_x, \partial_y\right)$.

In the limit that $\tau_f$ does not depend on $\eta_s$ we recover the Bjorken result
\begin{align}
  p^\alpha &d^3\Sigma_\alpha = \Big(m_T \cosh (y-\eta_s) - \vec{p}_T\cdot\vec{\nabla}_T\tau_f\Big)\tau_f dx dy d\eta_s\,.
\end{align}

In practice, we need to determine $d^3\Sigma_\alpha$ geometrically.
In previous works a simple algorithm has been used \cite{Hirano:2002ds,Hirano:2010pr} that adds a 
cuboidal volume element to the total freeze-out surface 
whenever the surface crosses a cell, e.g., if the quantity $\varepsilon-\varepsilon_{\rm FO}$ changes sign when moving along the $x$ direction,
one adds a volume element of size $\Delta y \Delta \eta_s \Delta \tau$ with its hyper-surface vector 
pointing in the $x$ direction (towards lower energy density).
So surface vectors always point along one of the axes $x$, $y$, $\eta_s$, or $\tau$.
This method overestimates the freeze-out surface itself but is sufficient for computing particle spectra.
However, it turns out that for computing anisotropic flow and especially higher harmonics than $v_2$ it is essential to determine
the freeze-out surface much more precisely. To do so, within \textsc{music} we employ the following method:

We define a cube in 4 dimensions that may reach over several lattice cells in every direction and over several $\tau$-steps,
and determine if and on which of the cube's 32 edges the freeze-out surface crosses. In this work we let the cube extend over one lattice cell
in each spatial dimension and over 10 steps in the time direction.
If the freeze-out surface crosses this cube, we use the intersection points to perform a 3D-triangulation of the three dimensional surface element 
embedded in four dimensional space.
This leads to a group of tetrahedra, each contributing a part to the hyper-surface-vector. This part is of the form
\begin{equation}\label{crossprod}
  d\Sigma^n_\mu = \varepsilon_{\mu\alpha\beta\gamma} A^\alpha B^\beta C^\gamma/6\,,
\end{equation}
where $A$, $B$, and $C$ are the three vectors that span the tetrahedron $n$. The factor $1/6$ normalizes the length of the vector
to the volume of the tetrahedron.
We demand that the resulting vector points into the direction of lower energy density, i.e., outwards. 
The vector-sum of the found tetrahedra determines the full surface-vector in the given hyper-cube.

Depending on where the freeze-out surface crosses the edges, the structure may be fairly simple (e.g. 8 crosses, all
on edges in $x$-direction) or rather involved (crossings on edges in many different directions).
The current algorithm is close to perfect and fails to construct hyper-surface elements only in very rare cases.
Typically these are cases when the surface crosses the cube in many different directions, e.g. in the $\eta_s$, $x$, and $\tau$ direction.
However, even for these cases a full reconstruction can usually be achieved and the algorithm was found to succeed 
in determining the volume element in $\sim 99\%$ of the cases for the studied systems. 
The $\sim 1\%$ of surface elements that could not be fully reconstructed usually miss only one tetrahedron. Because one typocally needs
between 8 and 20 tetrahedra to reconstruct a cell, the error introduced by missing one tetrahedron in the 1\% of the cells lies between 5 and 15\%.
Considering the high complexity of the triangulation procedure in four dimensions, this is a very satisfactory result.

\section{Results}
\label{results}
To obtain results for particle spectra, we first compute the thermal spectra of all particles and resonances up to 
$\sim 2\,{\rm GeV}$ using Eq.\,(\ref{cf}) and then perform resonance decays using routines 
from \textsc{azhydro} \cite{Sollfrank:1990qz,Sollfrank:1991xm,Kolb:2000sd,Kolb:2002ve}
that we generalized to three dimensions. 
Unless indicated otherwise, all shown results include the resonance feed-down.
Typically, the used time step size is $\Delta \tau\approx 0.01\,{\rm fm}/c$, and the spatial grid spacings
are $\Delta x=\Delta y= 0.08\,{\rm fm}$, and $\Delta \eta_s = 0.3$. This is significantly finer than in previous 
3+1D simulations: \cite{Hirano:2007ei} for example uses $\Delta \tau = 0.3\,{\rm fm}/c$,
$\Delta x=\Delta y= 0.3\,{\rm fm}$, and $\Delta \eta_s = 0.3$. The possibility to use such fine lattices 
is an improvement because it is mandatory when computing higher harmonics like $v_4$ as demonstrated below. 
Another advantage of using large lattices is that in the 
KT scheme the numerical viscosity decreases with increasingly fine lattices (see Appendix \ref{app:kt}).
The spatial extend of the lattice used in the following calcualtions is $20\,{\rm fm}$ in the $x$ and $y$ direction, and 20 units of
rapidity in the $\eta_s$ direction. 

\subsection{Particle spectra}
\begin{figure}[tb]
  \begin{center}
    \includegraphics[width=8.5cm]{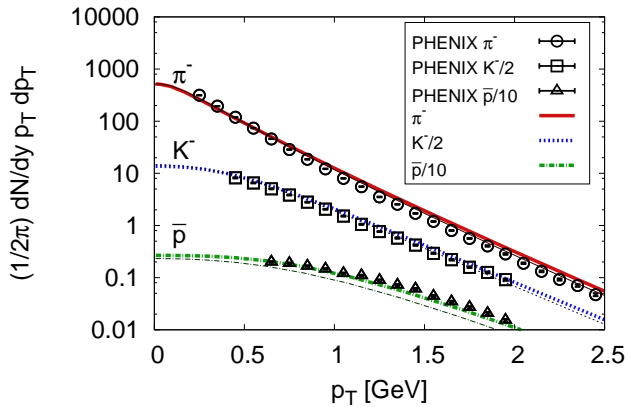}
    \caption{(Color online) $p_T$ spectra for $\pi^-$, $K^-$, and $\bar{p}$ at central 
      collisions using different equations of state (thin lines: AuAu-1 (EOS-Q), thick lines: AuAu-3 (EOS-L))
      compared to  0-5\% central PHENIX data \cite{Adler:2003cb}.
      The used impact parameter was $b=2.4\,{\rm fm}$.}
    \label{fig:AuAupt-}
  \end{center}
\end{figure}

In Fig.\,\ref{fig:AuAupt-} we present the transverse momentum spectra for identified particles in Au+Au collisions
at $\sqrt{s}=200\,{\rm GeV}$ compared to data from PHENIX \cite{Adler:2003cb}.
The used parameters are indicated in Table \ref{tab:parms}.
They were obtained by fitting the data at most central collisions.

We reproduce both pion and kaon spectra well. The model assumption of chemical equilibrium to very low temperatures leads to an
underestimation of the anti-proton spectrum. 
The overall shape is however well reproduced, even more so with the EOS-L that leads to flatter spectra \cite{Huovinen:2009yb}.

One way to improve the normalization of the proton and anti-proton spectra (as well as those of multistrange baryons)
is to employ the partial chemical equilibrium
model (PCE) \cite{Hirano:2002ds,Kolb:2002ve,Teaney:2002aj}, which introduces a chemical potential below a hadron species dependent 
chemical freeze-out temperature.
Note that the initial time was set to $\tau_0=0.4\,{\rm fm}/c$ when using the EOS-L to match the data. 
The quoted parameter sets fit the data very well, however,  they do not necessarily represent
the only way to reproduce the data and a more detailed anaylsis of the whole parameter space may find other
parameters to work just as well.

\begin{figure}[tb]
  \begin{center}
    \includegraphics[width=8.5cm]{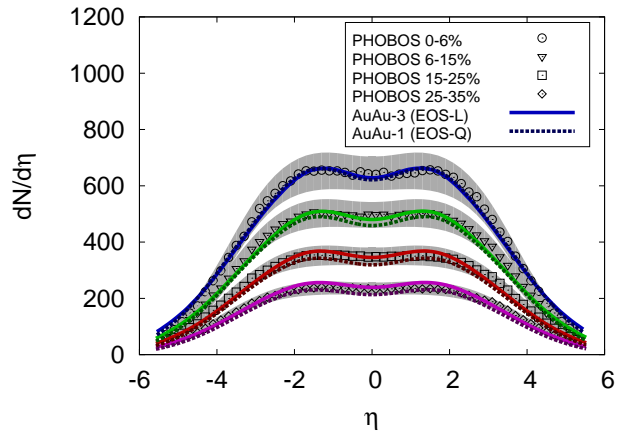}
    \caption{(Color online) Centrality dependence of pseudorapidity distribution compared to PHOBOS data \cite{Back:2002wb}.
    From top to bottom, the used average impact parameters are $b=2.4\,{\rm fm}$, $b=4.83\,{\rm fm}$, $b=6.7,{\rm fm}$, and $b=8.22\,{\rm fm}$.}
    \label{fig:AuAudNdeta}
  \end{center}
\end{figure}

\begin{figure}[tb]
  \begin{center}
    \includegraphics[width=8.5cm]{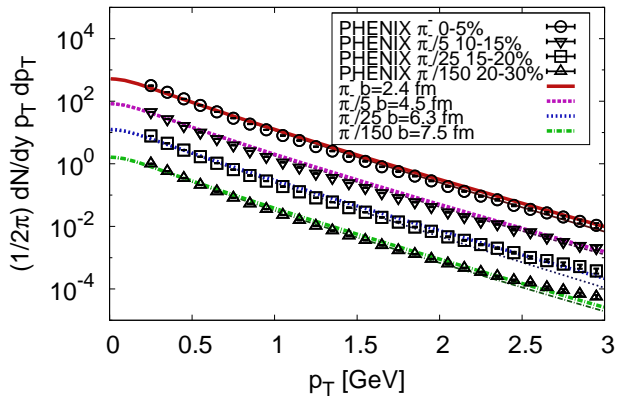}
    \caption{(Color online) Centrality dependence of $\pi^-$ transverse momentum spectra compared to PHENIX data  \cite{Adler:2003cb}.
    The curves (both data and hydro) for $10-15\,\%$, $15-20\,\%$ and $20-30\,\%$ centrality are scaled by a factor of 5, 25,and 150, respectively.
    Thick lines are for parameter set AuAu-3 (EOS-L), thin lines for AuAu-1 (EOS-Q).}
    \label{fig:AuAupipt}
  \end{center}
\end{figure}

Next, we show the pseudorapidity distribution of charged particles at different centralities compared 
to PHOBOS data \cite{Back:2002wb} in Fig.\,\ref{fig:AuAudNdeta}.
The only parameter that changes in going to larger centrality classes is the impact parameter. 
Experimental data is well reproduced also for semi-central collisions, showing that the results mostly depend on the collision geometry.
The used impact parameters, $b=2.4\,{\rm fm}$, $b=4.83\,{\rm fm}$, $b=6.7,{\rm fm}$, and $b=8.22\,{\rm fm}$, were obtained using the
optical Glauber model and correspond to the centrality classes used by PHOBOS. 
We show the centrality dependence of the transverse momentum spectrum of $\pi^-$ in Fig.\,\ref{fig:AuAupipt}.
Deviations occur for more peripheral collisions because the soft collective physics described by hydrodynamics 
becomes less important compared to jet physics in peri\-pheral events. 
However, we find smaller deviations than e.g. \cite{Nonaka:2006yn}.

\begin{center}
\begin{table*}
  \begin{tabular}{|c|c|c|c|c|c|c|c|c|c|}
    \hline
    set & EoS & $\tau_0 [{\rm fm}]$ & $\varepsilon_0 [{\rm GeV}/{\rm fm}^3]$ & $\rho_0 [1/{\rm fm}^3]$ 
    & $\varepsilon_{\rm FO} [{\rm GeV}/{\rm fm}^3]$ & $T_{\rm FO} [{\rm MeV}]$  
    & $\alpha$ & $\eta_{\rm flat}$ & $\sigma_{\eta}$ \\\hline
    AuAu-1 & EOS-Q & 0.55                & 41                               & 0.15                    & 0.09 & $\approx 130$
    & 0.25     & 5.9      & 0.4 \\
    AuAu-2 & EOS-Q & 0.55                & 35                               & 0.15                    & 0.09 & $\approx 130$
    & 0.05     & 6.0      & 0.3 \\
    AuAu-3 & EOS-L & 0.4                 & 55                               & 0.15                    & 0.12 & $\approx 137$
    & 0.05     & 5.9      & 0.4 \\
    \hline
  \end{tabular}
  \caption{Parameter sets. \label{tab:parms}}
\end{table*}
\end{center}

\begin{figure}[tb]
  \begin{center}
    \includegraphics[width=8.5cm]{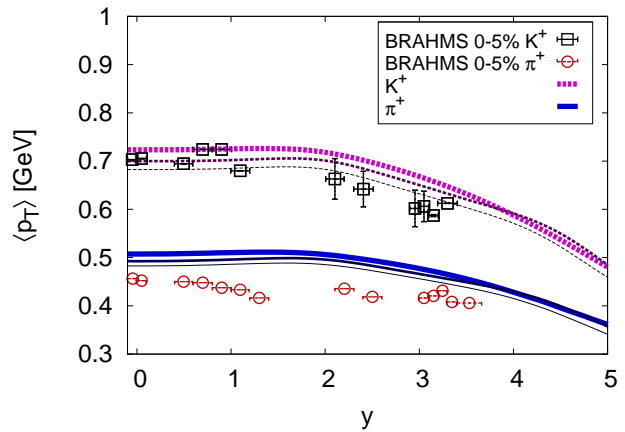}
    \caption{(Color online) $\langle p_T \rangle$ for positive kaons and pions 
      as a funciton of rapidity compared to most central BRAHMS data \cite{Bearden:2004yx}.
      The used impact parameter is $b=2.4\,{\rm fm}$. Different lines correspond to different parameter sets: From top to bottom:
      AuAu-3 (EOS-L), AuAu-1, AuAu-2 (EOS-Q).}
    \label{fig:avPTvsY}
  \end{center}
\end{figure}

Finally we present results for the average transverse momentum of pions and kaons as a function of pseudorapidity 
in central collisions. We compare
with $0-5\%$ central data by BRAHMS \cite{Bearden:2004yx} and find good agreement for kaons, but slightly larger values for pions.
This could be expected because the calculated $p_T$ spectra are slightly harder than the experimental data, especially when using
the EOS-L (see Fig.\,\ref{fig:AuAupt-}).  

\subsection{Elliptic flow}
\begin{figure}[tb]
  \begin{center}
    \includegraphics[width=8.5cm]{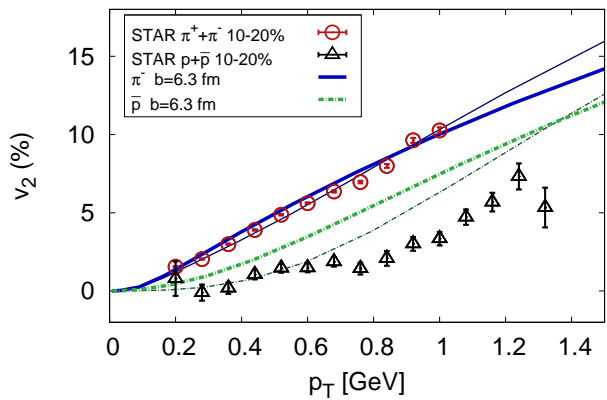}
    \caption{(Color online) $p_T$ dependence of the elliptic flow coefficient $v_2$ for $\pi^-$ and $\bar{p}$ 
      using parameter set AuAu-1 (EOS-Q, thin lines)
      and AuAu-3 (EOS-L, thick lines)
      compared to STAR data from \cite{Adams:2004bi}.}
    \label{fig:v2vspt}
  \end{center}
\end{figure}

We present results for $v_2$ as a function of $p_T$ integrated over the pseudorapidity range $-1.3<\eta<1.3$, which corresponds to 
the cut in the analysis by STAR \cite{Adams:2004bi} that we compare to.
We show results for identified hadrons obtained using parameter set AuAu-1 (EOS-Q) and AuAu-3 (EOS-L) in Fig.\,\ref{fig:v2vspt}.
While the pion elliptic flow is relatively well described for both equations of state, we
find an overestimation of the anti-proton $v_2$, especially when using the EOS-L. This is compatible with results in 
\cite{Huovinen:2009yb}. 
 
\begin{figure}[tb]
  \begin{center}
    \includegraphics[width=8.5cm]{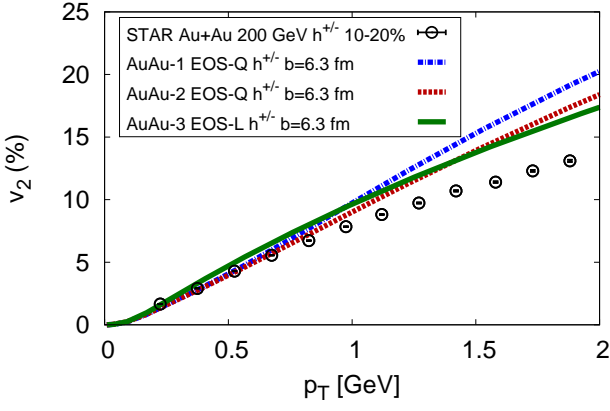}
    \caption{(Color online) $p_T$ dependence of the elliptic flow coefficient $v_2$ for charged hadrons 
      using parameter sets AuAu-1, AuAu-2 (EOS-Q), and AuAu-3 (EOS-L)
      compared to STAR data from \cite{Adams:2004bi}.}
    \label{fig:v2hvspt}
  \end{center}
\end{figure}

Charged hadron $v_2$ is presented in Fig.\,\ref{fig:v2hvspt} where we compare results using different contributions of binary collision 
scaling $\alpha$ which lead to different initial eccentricities. We also show the result obtained by using the EOS-L, which is 
somewhat above the EOS-Q result for lower $p_T$ but bends more strongly to be smaller at $p_T=2\,{\rm GeV}$.

Overall, we find that while the pion $v_2$ is well reproduced, both anti-proton and
charged hadron $v_2$ is overestimated for both parameter sets. So there is room for viscous corrections that have been 
found to reduce $v_2$ at $p_T=1.5\,{\rm GeV}$ by $20\,\%$ for $\eta_{\rm shear}/s=0.08$ 
\cite{Romatschke:2007mq,Dusling:2007gi,Molnar:2008xj,Song:2008si}.

\begin{figure}[tb]
  \begin{center}
    \includegraphics[width=8.5cm]{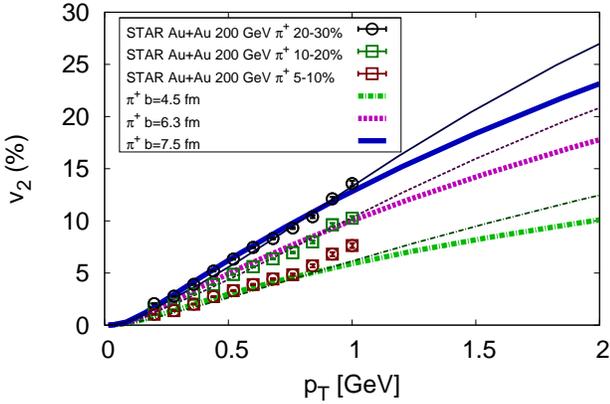}
    \caption{(Color online) Elliptic flow coefficient $v_2$ for positive pions using parameter set AuAu-1 (EOS-Q, thin lines) 
      and AuAu-3 (EOS-L, thick lines) for different centrality classes compared to STAR data from \cite{Adams:2004bi}.}
    \label{fig:v2vsptpi}
  \end{center}
\end{figure}

Fig.\,\ref{fig:v2vsptpi} shows $v_2$ of positive pions for different centrality classes, again comparing calculations using parameter sets
AuAu-1 and AuAu-3 with experimental data \cite{Adams:2004bi}. 
In both cases the agreement with the experimental data that is available to up to $p_T=1\,{\rm GeV}$ is very reasonable. 

\begin{figure}[tb]
  \begin{center}
    \includegraphics[width=8.5cm]{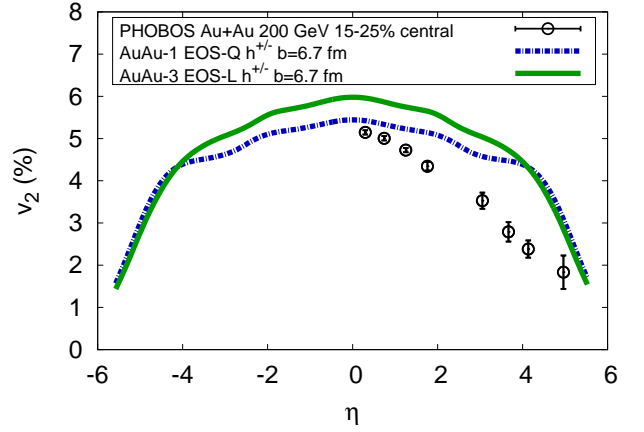}
    \caption{(Color online) Pseudorapidity dependence of the elliptic flow coefficient $v_2$ for charged hadrons using parameter sets 
      AuAu-1 (EOS-Q), and AuAu-3 (EOS-L) compared to PHOBOS data from \cite{Back:2004mh}.}
    \label{fig:v2vseta}
  \end{center}
\end{figure}

In Fig.\,\ref{fig:v2vseta} we present the result for $v_2$ as a function of pseudorapidity $\eta$, comparing to data from PHOBOS \cite{Back:2004mh}. 
As earlier calculations 
\cite{Hirano:2002ds,Nonaka:2006yn} the hydrodynamic model calculation overestimates the elliptic flow especially at forward and
backward rapidities. This is most likely due to the fact that the assumption of ideal fluid behavior is no longer valid
far away from the midrapidity region. Calculations combining hydrodynamic evolution with a hadronic after-burner improve on this 
\cite{Hirano:2005xf,Nonaka:2006yn}. 
Effects of viscosity on $v_2(\eta)$ in the 3+1 dimensional simulation have been estimated to be stronger at larger $|\eta|$ 
\cite{Bozek:2009mz} and it will be interesting to see what a full computation will yield.

\subsection{Higher harmonics}
The extraction of higher harmonic coefficients from the computed particle distributions has to be done with great care.  
Apart from being highly sensitive to the initial conditions \cite{Kolb:2003zi,Huovinen:2005gy}, the fourth harmonic coefficient $v_4$
is also highly sensitive to the discretization of the freeze-out surface and lattice artifacts. 
Where other quantities such as $p_T$ spectra and $v_2$
are almost unaffected by a change of the lattice resolution or the freeze-out method, $v_4$ depends strongly on the method and 
the lattice spacing. Using the simplified freeze-out surface algorithm described above, the dependence of $v_4$ on the discretization 
becomes very strong (in this case $v_4$ is negative when using a $128^3$-lattice 
and only becomes positive and slowly approaches the correct value for much finer lattices).

\begin{figure}[tb]
  \begin{center}
    \includegraphics[width=8.5cm]{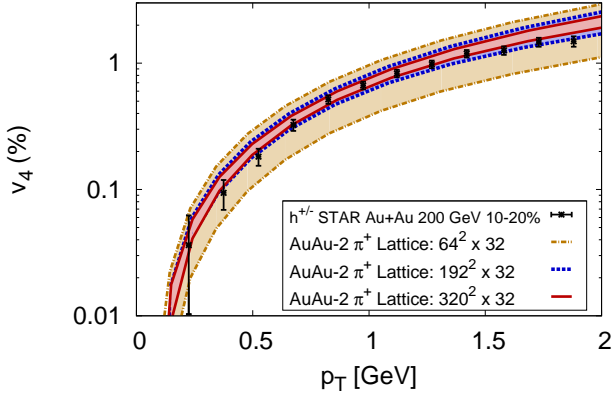}
    \caption{(Color online) Pion-$v_4$ (without resonance decays) as a function of $p_T$ computed on different lattices.
      The upper curve of each band is the result from when the impact parameter is aligned with the diagonal in the $x$-$y$-plane,
      the lower curve from when it is aligned with the $x$-axis. The absolute value of the impact parameter is $b=6.3\,{\rm fm}$.
      We compare to charged hadron STAR data from \cite{Adams:2004bi}.}
    \label{fig:v4error}
  \end{center}
\end{figure}

It is therefore necessary to work on very fine lattices and have a very sophisticated algorithm for determining the freeze-out surface
in order to obtain reliable results for $v_4$.
To measure the error introduced by the anisotropic discretization of the lattice (lattice along the diagonal in the transverse plane
looks different than along one of the axes), we compute $v_4$ twice: once with the impact paramter along the $x$-axis, once
with the impact parameter along the diagonal in the $x$-$y$-plane. The difference between the results is a measure 
of discretization errors in $v_4$ and is shown for the pion $v_2$ in Fig.\,\ref{fig:v4error}. 
The difference decreases significantly when going 
from a $64^2$ to a $320^2$ lattice in the transverse plane. Hence, the numerical error of $v_4$ is well under control.  

\begin{figure}[tb]
  \begin{center}
    \includegraphics[width=8.5cm]{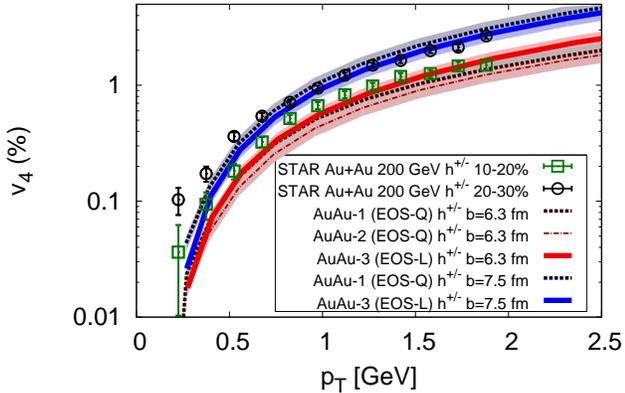}
    \caption{(Color online) $v_4$ for charged hadrons using all parameter sets
      for different centrality classes compared to STAR data from \cite{Adams:2004bi}. See text for details.}
    \label{fig:v2vspth}
  \end{center}
\end{figure}

\begin{figure}[tb]
  \begin{center}
    \includegraphics[width=8.5cm]{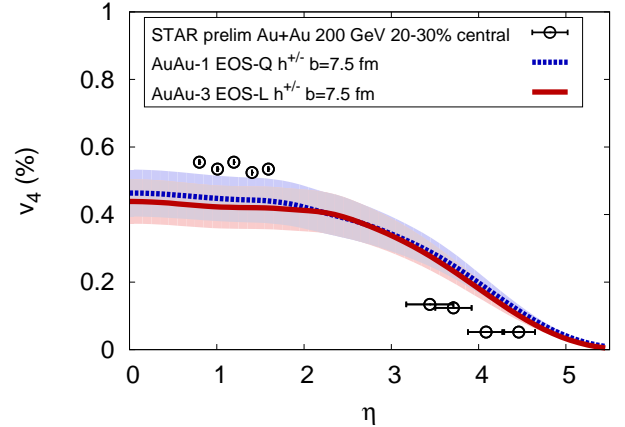}
    \caption{(Color online) $v_4$ for charged hadrons using parameter sets AuAu-1 and AuAu-3 at $b=7.5\,{\rm fm}$
     compared to STAR data from \cite{Bai:2007ky}.}
    \label{fig:v4vseta}
  \end{center}
\end{figure}

Fig.\,\ref{fig:v2vspth} shows $v_4$ of charged hadrons computed with both parameter set AuAu-1 (EOS-Q) and AuAu-3 (EOS-L).
We added error bands representing an estimate for the discretization error on the used $256^2\times 64$ lattice.
Motivated by the results shown in Fig.\,\ref{fig:v4error}, we choose $\pm 15\%$.
Experimental data for mid-central centrality classes is well reproduced in both cases, and 
contrary to expectations \cite{Kolb:2003zi}, 
we find that $v_4$ is not very sensitive to either the initial condition or the equation of state.
This is also visible in Fig.\,\ref{fig:v4vseta} where we show $v_4$ as a function of pseudorapidity compared to preliminary 
STAR data \cite{Bai:2007ky}. Also here we add an error band to indicate the discretization error, estimating it to be $\pm 15\%$.

We have checked that the ratio $v_4/v_2^2$ approaches $0.5$ for large $p_T$
as it should for an ideal fluid, at least in the limit of small impact parameter \cite{Borghini:2005kd}.
The difference to the data, which for charged hadrons in minimum bias collisions is about constant at $v_4/v_2^2\approx 1.2$ 
\cite{Poskanzer:2004vd,Adams:2004bi} comes mostly from our overestimation of $v_2$ at high $p_T$, while $v_4$ is well reproduced 
(see Figs.\,\ref{fig:v4error} and \ref{fig:v2vspth}).



\section{Conclusions and Outlook}
\label{conclusions}
We presented first results from our newly developed 3+1 dimensional relativistic fluid dynamic simulation, \textsc{music}, using the
Kurganov-Tadmor high-resolution central scheme to solve the hydrodynamic equations. The method handles large gradients very well, which
makes it ideal for future explorations of `lumpy' initial conditions or energy-momentum deposition by jets.
It also has a very small numerical viscosity, which is a prerequisite for extracting physical viscosities in the future extension to 
dissipative hydrodynamics.

We showed a detailed comparison of results using different equations of state including a very recent parametrization
of combined lattice and hardon resonance gas equations of state. Our calculations of identified hadron $p_T$-spectra, pseudorapidity
distributions and elliptic flow coefficients of charged hadrons in Au+Au collisions at the highest RHIC energies
reproduced results of earlier 3+1 dimensional simulations. 
In addition, we were able to obtain reliable results for the anisotropic flow coefficient $v_4$, which is highly 
sensitive to discretization errors. 
This was possible by developing a sophisticated algorithm for determining the freeze-out surface in four dimensions 
and running the simulation on fine lattices on many processors parallely to obtain results within a reasonable amount of time.
We found that contrary to earlier expectations, $v_4$ is not very sensitive to the initial conditions. Neither is it very sensitive 
to the equation of state. 

The next step will be the inclusion of viscous effects which we will present in a forthcoming work.
We also plan to combine the simulation with our event generator for the hard probes, \textsc{martini},
to finally obtain a coupled simulation of both the soft and hard physics in heavy-ion collisions,
creating an unprecedented theoretical tool for the study of the hot and dense phase of matter generated in heavy-ion collisions. 

\section*{Acknowledgments}
\hyphenation{Pasi Huovinen Abhijit Majumder Ulrich Heinz Michael Strickland Richard Tomlinson Steffen Bass Evan Frodermann
Chiho Nonaka Huichao Song Guy Moore Guang Qin Vasile Topor-Pop}
We are happy to thank Steffen Bass, Kevin Dusling, Evan Frodermann, Ulrich Heinz, Tetsufumi Hirano, Pasi Huovinen, 
Chiho Nonaka, Paul Romatschke, Huichao Song, Derek Teaney, and Raju Venugopalan for very useful discussions. 

This work was supported in part by the Natural Sciences and Engineering Research Council of Canada. 
B.S.\ gratefully acknowledges a Richard H.~Tomlinson Postdoctoral Fellowship by McGill University.

\appendix
\section{Kurganov-Tadmor Method}
\label{app:kt}
As described in the main text, the Kurganov-Tadmor method is a MUSCL-type
finite volume method in which the cell average of the density $\rho$ around $x_j$ is used
instead of the value of the density at $x_j$.
To do so, we first divide the space into equal intervals of the width 
$\Delta x$.
If one integrates over the interval $[x_j-\Delta x/2, x_j+\Delta x/2]$,
the conservation equation becomes
\begin{align}
{d\over dt} \int_{x_{j-1/2}}^{x_{j+1/2}} dx\,&\rho(x,t) = \nonumber \\
& {J(x_{j-1/2},t) - J(x_{j+1/2},t)}
\label{eq:discrete_v}
\end{align}
where we introduced the notations $x_{j\pm 1/2} = x_j \pm \Delta x/2$.
Defining the cell average at $x_j$ as
\be
\barrho_j(t) = 
{1\over \Delta x} \int_{x_{j-1/2}}^{x_{j+1/2}} dx\, \rho(x,t)\,,
\ee
the above equation becomes
\be
{d\over dt}\barrho_j(t) = {J(x_{j-1/2},t) - J(x_{j+1/2},t)\over \Delta x}\,. 
\label{eq:finite_v2}
\ee
Using this exact equation, we can formally advance the time by
$\Delta t$ as
\begin{align}
&\barrho_j(t+\Delta t)
=\barrho_j(t)
\nonumber\\
&\qquad -{1\over \Delta x} 
\int_{t}^{t+\Delta t} dt'\left(
J(x_{j+1/2},t') - J(x_{j-1/2},t') 
\right)\,.
\label{eq:volume_ave}
\end{align}
So far no approximation has been made. 
The main variables to calculate are the discrete average values
$\barrho_j^n \equiv \barrho_j(t_n)$ for all $x_j = x_0 + j\Delta x$
and at every time step $t_n = t_0 + n\Delta t$.
Note that in Eq.\,(\ref{eq:volume_ave}), the currents are evaluated at
half-way points between $x_j$ and its neighboring points $x_{j\pm 1}$.
Therefore, even if
Eq.\,(\ref{eq:volume_ave}) is exact, it is not complete. One needs to know how
to evaluate $J(x_{j\pm 1/2}, t')$ which in turn needs the value of 
$\rho(x_{j\pm 1/2}, t')$. Note further that this 
is {\em not} the cell averages but actual local values of $\rho$. 
Thus, the problem to solve now is how to approximate the current and the charge
density at arbitrary $x$ from the cell averages at discrete points $x_j$.

A simple but effective solution is 
to approximate the local value with the average value
and make a linear interpolation within each cell:
\begin{align}
p(x,t_n) &= \sum_j \left[\barrho_j^n + (\rho_x)^n_j(x-x_j)\right]\nonumber\\
&\qquad\qquad\times\theta(x_{j-1/2} < x < x_{j+1/2})
\label{eq:linear_approx}
\end{align}
where $\theta(x_{j-1/2}\le x \le x_{j+1/2})$ is defined to be 1 when the
condition is fulfilled and 0 otherwise.
This piecewise
linear approximation is constructed in such a way that the amount of
matter in the cell remains the same, that is, by construction
\be
\int_{x_{j-1/2}}^{x_{j+1/2}}dx\, p(x, t_n)
=
\barrho_j^n
\ee
which, of course is a necessary condition for solving a conservation equation.
The derivative $(\rho_x)^n_j$ is a suitable approximation of
$\partial_x\rho$ at $x_j$ and $t_n$ constructed from the cell averages.
It could be the backward slope
\be
(\rho_x)^n_j = {\barrho_j^n - \barrho_{j-1}^n\over \Delta x}\,,
\label{eq:backward}
\ee
or the forward slope
\be
(\rho_x)^n_j = {\barrho_{j+1}^n - \barrho_{j}^n\over \Delta x}\,,
\label{eq:forward}
\ee
or any combination of them.
We will discuss the choice of derivatives in more detail shortly.
For now we leave this choice open.

There is, however, a potentially serious problem with this piecewise linear
reconstruction.
There are two ways to calculate the value of 
$p(x_{j+1/2},t^n) \equiv p_{j+1/2}^n$.
One way is to calculate if from the left
\be
\left. p_{j+1/2}^n \right|_{\rm left} 
= \barrho_j^n + (\rho_x)_j^n \Delta x/2\,,
\ee
or from the right
\be
\left. p_{j+1/2}^n \right|_{\rm right}
= \barrho_{j+1}^n - (\rho_x)_{j+1}^n \Delta x/2\,.
\ee
These two values in general do not coincide.
Since we need these half-way
values to calculate the currents at the cell boundaries,
we need to find a way to deal with this discontinuity.

A simple but effective solution
to this problem was originally proposed by Nessyahu and
Tadmor \cite{Nessyahu:1990}: First, the discontinuity matters only for the current
part because of the spatial derivative. For the density part, 
the linear approximation is fine because only the time derivative is taken.
For the density, we can integrate over this discontinuity without any problem.
Now, for a sufficiently small time interval $\Delta t$, the effect of the
discontinuity at $x_{j+1/2}$ will not reach $x_{j}$ and $x_{j+1}$.
Hence, if one alternatively
considers a cell defined by $[x_j, x_{j+1}]$ instead of one
defined by $[x_{j-1/2}, x_{j+1/2}]$, then the currents are calculated at
$x_j$ and $x_{j+1}$ where $p(x,t)$ is still smooth.
Hence we can use $p(x,t)$ to calculate both 
the charge density and the currents
on the right hand side of Eq.\,(\ref{eq:volume_ave}) provided that the average
is now over the staggered cell $[x_j, x_{j+1}]$. 
After this step, the discontinuities are
located at $x_j$'s instead of the half-way points.
The next step in this approach is to repeat the same procedure now for
$[x_{j-1/2},x_{j+1/2}]$ where this time the half-way points
$x_{j\pm 1/2}$ are where
the linear interpolation is smooth. This staggered grid approach
is an effective method. However, the numerical viscosity term turns out to
be $O((\Delta x)^4/\Delta t)$ which means that one still cannot take the
$\Delta t\to 0$ limit. 

Generalizing the Nessyahu-Tadmor method, Kurganov and Tadmor came up with a
better solution to this problem.  Their main idea can be described as
follows: The Nessyahu-Tadmor method relies on the smallness of $\Delta t$ to
guarantee that the influence of the discontinuities does not reach the
mid-points of the (staggered) cells. This can be further improved if one
uses one more piece of information, namely the maximum local propagation speed.
The influence of the discontinuities at $x_{j\pm 1/2}$ 
can travel no faster than the maximum propagation speed
given by $a = \left| \partial J/\partial \rho \right|$.
Therefore, it makes sense to divide the space into two groups.
One such group is given by the following set of intervals 
\begin{equation}
\mu_{j+1/2}^n 
= [x_{j+1/2}-a_{j+1/2}^n\Delta t, x_{j+1/2}+a_{j+1/2}^n \Delta t]\,,
\end{equation}
where $a_{j+1/2}^n$ is the maximum propagation speed at $x_{j+1/2}$ and time
$t_n$.
The linear interpolation $p(x,t)$
is possibly discontinuous in $\mu_{j+1/2}^n$
as indicated by the $\theta$-functions in Eq.\,(\ref{eq:linear_approx}).
The other group is given by the set
\begin{equation}
\chi_{j}^n = [x_{j-1/2}+a_{j-1/2}^n \Delta t, x_{j+1/2}-a_{j+1/2}^n \Delta t]\,,
\end{equation}
and in these intervals, $p(x,t)$ is linear and smooth.
The fact that we must have non-empty $\chi_{j}^n$
gives us a condition on $\Delta t$
\be
\Delta t < {\Delta x\over a_{j+1/2}^n + a_{j-1/2}^n}\,,
\ee
which is related to the 
CFL (Courant-Friedrichs-Lewy) condition.
In fact, Ref.\,\cite{Kurganov:2000} has a more severe CFL condition
\be
\Delta t \le {\Delta x\over 8 a_{\rm max}} \,,
\ee
where $a_{\rm max}$ is the maximum propagation speed in the whole region.

The derivation of the Kurganov-Tadmor method now proceeds as follows. First,
apply Eq.\,(\ref{eq:volume_ave}) to $\mu_{j\pm 1/2}^n$ and $\chi_j^n$.
Since we have divided the space into the $\mu$-set and the $\chi$-set, 
the currents are not being evaluated at the discontinuities at $x_{j\pm 1/2}$.
Hence we can safely use $p(x,t)$ to evaluate
the right hand side of Eq.\,(\ref{eq:volume_ave}) in both
$\mu^n_{j\pm 1/2}$ and $\chi_j^n$. In this way, we get estimates of
the density average in these intervals at the next time step.
The intervals $\mu_{j\pm 1/2}^n$'s 
and $\chi_j^n$'s are non-uniformly distributed.
However, our starting point was the cell averaged values in the uniform
grid.
The next step is then to project the next-time values in
this non-uniform grid of $\mu_{j\pm 1/2}$'s and $\chi_j^n$'s 
onto the original uniform grid $[x_{j-1/2}, x_{j+1/2}]$.
Finally, one takes the $\Delta t\to 0$ limit to get the semi-discrete
equations. 

Application of Eq.\,(\ref{eq:volume_ave}) to $\mu_{j\pm 1/2}^n$ and $\chi_j^n$
proceeds as follows.
Within $\mu_{j+1/2}^n$, the right hand side of Eq.\,(\ref{eq:volume_ave})
is given by
\begin{widetext}
\begin{equation}
w^{n+1}_{j+1/2}
\equiv 
{1\over 2a_{j+1/2}^n\Delta t}
\int_{\mu_{j+1/2}^n}\hspace{-0.8cm} 
d\xi\, \rho(\xi, t_{n})
-
{1\over 2a_{j+1/2}^n\Delta t}
\int_{t_n}^{t_{n+1}}\hspace{-0.6cm} dt'\,
\Big(
J(x_{j+1/2}+a_{j+1/2}^n \Delta t, t)
-
J(x_{j+1/2}-a_{j+1/2}^n \Delta t, t)
\Big)
\end{equation}
\end{widetext}
where $2a_{j+1/2}^n\Delta t$ is the size of the interval.
Using $p(x,t)$ for the $\rho$ integral and using the
mid-point rule for the time integral, we obtain
\be
w^{n+1}_{j+1/2} \hspace{-0.1cm}
&=&\hspace{-0.1cm}
{\barrho_j^n + \barrho_{j+1}^n\over 2}
+
{\Delta x - a_{j+1/2}^n\Delta t\over 4}
\left( (\rho_x)^n_j - (\rho_x)^n_{j+1} \right)
\non & & {}
-{1\over 2a_{j+1/2}^n}
\Big(
J(x_{j+1/2,+}^n, t+\Delta t/2)\non & & {}
\hspace{2cm}
-
J(x_{j+1/2,-}^n, t+\Delta t/2)
\Big)\,,
\ee
where we have defined
\be
x_{j + 1/2,\pm}^n = x_{j+1/2}\pm a_{j+1/2}^n\Delta t\,.
\ee
Similarly,
for $\chi_j^n$, 
\be
w_j^{n+1}
&\equiv &
\barrho_j^n 
-
{1\over 2}
(\rho_x)^n_j
(a_{j+1/2}^n - a_{j-1/2}^n)\Delta t
\non & & {}
-
{(\Delta t/\Delta x) \over 1-(\Delta t/\Delta x) (a_{j-1/2}^n + a_{j+1/2}^n)}\non & & {}
\times
\Big(
J(x_{j+1/2,-}^n, t+\Delta t/2)\non & & {}
\hspace{0.6cm}
-
J(x_{j-1/2,+}^n, t+\Delta t/2)
\Big)\,.
\ee

At this point, $w^{n+1}_j$ and $w^{n+1}_{j+1/2}$ approximate the value of
$\barrho$ at $t_{n+1}$ on a non-uniform grid. 
The next step is to construct a piecewise linear
function $q(x,t_{n+1})$ using $w^{n+1}_j$ and $w^{n+1}_{j+1/2}$
and integrate over $[x_{j-1/2}, x_{j+1/2}]$ to
get the next cell average $\barrho_j^{n+1}$. 
For this purpose,
the piecewise linear function $q(x, t_{n+1})$ is constructed by using
a linear approximation within $\mu^n_{j+1/2}$ and the constant approximation
within $\chi_j^n$
\be
q(x, t^{n+1}) &=&  \non & & {}\hspace{-1.5cm}
\sum_j
\Big\{
[w^{n+1}_{j+1/2} + (\rho_x)^{n+1}_{j+1/2}(x-x_{j+1/2})] \non & & {}
\hspace{-1cm}
\times
\theta(x\in \mu_{j+1/2}^n) + w_j^{n+1}\theta(x \in \chi_j^n)
\Big\}
\ee
The derivative appearing in the above approximation must also be calculated
using $w_{j+1/2}^{n+1}$ and $w_j^{n+1}$.
Again leaving what to use for the derivate
for later discussions, integrating
over the interval $(x_{j-1/2}, x_{j+1/2})$ 
finally yields the cell average at the next time step
\be
\barrho_j^{n+1} 
& = & 
w_{j-1/2}^{n+1}a_{j-1/2}^n {\Delta t\over \Delta x}+w_{j+1/2}^{n+1}a_{j+1/2}^n {\Delta t\over \Delta x}
\non & & {}
+\left(1 - \frac{\Delta t}{\Delta x}(a_{j+1/2}^n+a_{j-1/2}^n)\right) w_j^{n+1}
\non & & {}
+ (\rho_x)^{n+1}_{j-1/2}{(a_{j-1/2}^n\Delta t)^2\over 2\Delta x}\non & & {}
- (\rho_x)^{n+1}_{j+1/2}{(a_{j+1/2}^n\Delta t)^2\over 2\Delta x}\,.
\ee

Passing to the $\Delta t\to 0$ limit, we get Kurganov and Tadmor's main
result in the semi-discrete form
\be
{d\over dt}\barrho_j(t) = -{H_{j+1/2}(t) - H_{j-1/2}(t)\over \Delta x}\,,
\ee
where 
\be
H_{j\pm 1/2} 
&=&
{J(x_{j\pm 1/2,+},t) + J(x_{j\pm 1/2,-},t)\over 2}\non & & {}
\hspace{-1cm}-
{a_{j\pm 1/2}(t)\over 2}
\left(
\barrho_{j\pm 1/2,+}(t) - \barrho_{j\pm 1/2,-}(t)
\right)
\ee
Here 
\be
\barrho_{j+1/2,+} &=& \barrho_{j+1}-{\Delta x\over 2}(\rho_x)_{j+1}\,,
\\
\barrho_{j+1/2,-} &=& \barrho_j + {\Delta x\over 2}(\rho_x)_j\,,
\ee
and $J(x_{j\pm 1/2,\pm})$ are evaluated with $\barrho_{j+1/2,\pm}$.
Any explicit $x$ dependence in $\barrho_{j+1/2,\pm}$ and
$J(x_{j\pm 1/2,\pm})$ must be evaluated at $x_{j+1/2}$.
Note that all references to the intermediate values have disappeared.

One detail we need to take care of now is the choice of the spatial
derivatives. 
Formally, the second-order approximation
\be
(\rho_x) \approx {\rho_{j+1}-\rho_{j-1}\over 2\Delta x}
\ee
gives a better approximation than the first-order approximations
Eq.\,(\ref{eq:backward}, \ref{eq:forward}).
But it is also known that when there is a stiff gradient, the second order
expression tends to introduce spurious oscillations in the solution. 
To remedy this situation, one needs to use flux limiters which automatically
switch the form of the numerical derivative according to the stiffness of
the slope.
Kurganov and Tadmor chose the minmod limiter given by
\begin{equation}
(\rho_x)_j
=
\hbox{minmod}
\left(
\theta {\barrho_{j+1}-\barrho_j\over \Delta x}, 
{\barrho_{j+1}-\barrho_{j-1}\over 2\Delta x},
\theta {\barrho_{j}-\barrho_{j-1}\over \Delta x}
\right)\nonumber
\end{equation}
where
\be
\hbox{minmod}(x_1, x_2, \cdots) = \left\{
\begin{array}{ll}
\hbox{min}_j\{x_j\},& \hbox{if $x_j> 0$ $\forall j$}\\
\hbox{max}_j\{x_j\},& \hbox{if $x_j< 0$ $\forall j$}\\
0, & \hbox{otherwise}\nonumber
\end{array}
\right.
\ee
and $1\le \theta \le 2$ is a parameter that controls the amount of
diffusion and the oscillatory behavior. This is also our choice with $\theta
= 1.1$.

\bibliography{hydroNoEprint}

\end{document}